\documentclass[aps,10pt,prd,groupedaddress,nofootinbib,notitlepage,eqsecnum,preprintnumbers]{revtex4-1}

\usepackage[utf8]{inputenc}
\usepackage{hyperref}
\usepackage{xcolor}
\usepackage{graphicx}
\usepackage{amsmath,amssymb,amsbsy,amstext,amsthm}
\usepackage{mathtools}
\usepackage{amsfonts}
\usepackage{comment}

\newenvironment{equations}{\equation\aligned}{\endaligned\endequation}

\def\beq{\begin{equation}}
\def\eeq{\end{equation}}
\def\ba{\begin{equations}}
	\def\ea{\end{equations}}
\def\bc{\begin{center}}
	\def\ec{\end{center}}

\def\n{\nabla}

\def\pa{\partial}

\definecolor{DarkBlue}{rgb}{0,0,0.7}

\begin{document}
	\title{{{\boldmath
		Could the black hole singularity be a field singularity?}}}
	
		\author{\textsc{Guillem Dom\`enech$^{a}$}}
		\email{{domenech}@{thphys.uni-heidelberg.de}}
		
		\author{\textsc{Atsushi Naruko$^{b, c}$} }
		\email{{naruko}@{yukawa.kyoto-u.ac.jp}}

		\author{\textsc{Misao Sasaki$^{b,d,e,f}$} }
		\email{{misao.sasaki}@{ipmu.jp}}
	
		\author{\textsc{Christof Wetterich$^{a}$} }
		\email{{c.wetterich}@{thphys.uni-heidelberg.de}}
	
		\affiliation{
			$^{a}$\small{Institut f\"ur Theoretische Physik, Ruprecht-Karls-Universit\"at Heidelberg, Philosophenweg 16, 69120 Heidelberg, Germany}\\
			$^{b}$\small{Center for Gravitational Physics, Yukawa Institute for Theoretical Physics, Kyoto University, Kyoto 606-8502, Japan}\\
			$^{c}$\small{Frontier Research Institute for Interdisciplinary Sciences \& Department of Physics, Tohoku University, 
Sendai 980-8578, Japan}\\
			$^{d}$\small{Kavli Institute for the Physics and Mathematics of the Universe (WPI), Chiba 277-8583, Japan}\\
			$^{e}$\small{CAS Key Laboratory of Theoretical Physics, Institute of Theoretical Physics, Chinese Academy of Sciences, Beijing 100190, China}\\
			$^{f}$\small{Leung Center for Cosmology and Particle Astrophysics, National Taiwan University, Taipei 10617}
		}
	\date{\today}

\preprint{TU1084,YITP-19-53}
	
\begin{abstract} 
In the wake of interest to find black hole solutions with scalar hair, we investigate the effects of disformal transformations on static spherically symmetric space-times with a non-trivial scalar field. In particular, we study solutions that have a singularity in a given frame, while the action is regular. We ask if there exists a different choice of field variables such that the geometry and the fields are regular. 
We find that in some cases disformal transformations can remove a singularity from the geometry or introduce a new horizon.
This is possible since the Weyl tensor is not invariant under a general disformal transformation. There exists a class of metrics which can be brought to Minkowksi geometry by a disformal transformation, which may be called disformally flat metrics. We investigate three concrete examples from massless scalar fields to Horndeski theory for which the singularity can be removed from the geometry. This might indicate that no physical singularity is present. We also propose a disformal invariant tensor.
\end{abstract}

	\maketitle

\section{Introduction\label{sec:introduction}}
When we speak of the universe expanding or shrinking this means that a given field, the metric field, changes its value in time. There is no ``creation'' or ``destruction'' of space -- only a field value changes. Since this field value influences the propagation of photons we can associate it with geometrical properties. In quantum field theory observables are expressed in terms of fields, but the choice of fields is not unique. Field transformations do not change the predictions of observables if the expressions of observables in terms of fields are transformed accordingly.

Field transformations that change the metric are called ``frame-changes''. On the level of the quantum field equations derived from the quantum effective action all the frames are fully equivalent \cite{Wetterich:2014gaa} as long as the frames are related by invertible transformations. Frame-changes are simply a change of variables in differential equations. Nevertheless, frame changes can affect strongly the geometry. Geometries with singularities in one frame may turn out completely regular in another frame. The freedom in the choice of frames may be called ``field relativity''. This extended notion of relativity goes beyond general relativity since frame changes are, in general, not coordinate transformations.

It can happen that regular field configurations are mapped to geometries with a singular geometry by a singular field transformation. In this case there is no physical singularity. The singularity is solely a ``field singularity'', due to a perhaps inconvenient choice of fields. This is somewhat analogous to coordinate singularities. It has been argued that the big bang singularity in models of inflationary cosmology is only a field singularity \cite{Wetterich:2014gaa} (see also \cite{Bahamonde:2016wmz}). In this paper we address the question if it could be possible that the central singularity of black holes is a field singularity that can be removed by the choice of a different frame for the metric. Previous studies of the central singularity in scalar-tensor theories showed that conformal transformations may relate a naked singularity in the Einstein frame and a wormhole in the Jordan frame \cite{Faraoni:2015paa,Faraoni:2016ozb}. However, since the black hole singularity is in the squared Weyl tensor, it can in general not be removed solely by conformal transformations. We therefore investigate the role of disformal field transformations. Since the wanted frame change involves necessarily other fields, typically scalar fields, the outcome of this investigation depends strongly on the field content of models beyond the purely metric sector.

Singularities in General Relativity  are present under a few assumptions \cite{PhysRevLett.14.57,Hawking:1969sw} (for a review see Ref.~\cite{Hawking2014}). Well known examples are the central singularity in black holes (\textit{e.g.} Schwarzschild metric) and the initial singularity in cosmology (\textit{e.g.} Friedman-Lema\^{i}tre-Robertson-Walker metric). Such singularities are often regarded as ``physical'' if they yield a geodesically incomplete manifold, for which time-like geodesics start/end abruptly after a finite proper time \cite{Borde:2001nh,Hawking2014}. Proper time is, however, a frame dependent quantity \cite{Wetterich:2014zta} and the issue of geodesic incompleteness has to be discussed with care.   
Being an undesirable feature of a theory, there might be mechanisms to seal singularities from an observer. For example, one could propose that $(i)$ singularities are hidden by horizons (Cosmic Censorship Conjecture \cite{1969NCimR...1..252P,Wald:1997wa}) or that $(ii)$ the theory is modified such that singularities are not allowed, \textit{e.g.}, bouncing universes \cite{Qiu:2011cy}, limiting curvature mechanisms \cite{1982JETPL..36..265M,1988ZhETF..94....1G,Frolov:1988vj} (for recent studies see Refs.~\cite{Yoshida:2017swb,BenAchour:2017ivq}), usually implying that the weak energy condition is violated (see Ref.~\cite{Rubakov:2014jja} for a review). However, metrics with a singularity are typically solutions of a simplified model, \textit{e.g.} the matter content is absent, is a scalar field or a perfect fluid, and assumes that matter is minimally coupled to the metric. Non-minimal couplings might change the conclusions of the simplified model and, perhaps, yield a manifold without singularities.

Another question that raises is that even though the solution is singular, if the action is regular then the solution might be physically acceptable. This was proposed by Hawking and Turok for the creation of an open universe through a singular instanton \cite{Hawking:1998bn}. Could that mean that if the action is regular, there exists a transformation of the metric and the fields that yields a regular solution? For example, the Hawking-Turok instanton is regular when embedded in one higher dimension as was shown by Garriga \cite{Garriga:1998ri}. Furthermore, note that this transformation of the metric and the fields, usually referred as changing ``frames'', can be interpreted as matter being not minimally coupled, in line with the discussion in the previous paragraph.

If we focus for the moment on pre-inflationary cosmology, it is well known that there are frames for which there is no initial singularity in the metric \cite{Wetterich:2013jsa,Wetterich:2013aca}. The reason is that a Friedmann-Lema{\^i}tre-Robertson-Walker (FLRW) metric is conformally flat and the singularity can be removed from the metric by a conformal transformation. For typical inflationary models there are frames for which the initial singularity is postponed to past infinity in the physical time. In this case the singularity has not been removed from the solution, as one would expect from a regular change of variables. There exists also particular ``primordial flat frames''  for which spacetime becomes flat in the infinite past \cite{Wetterich:2014gaa}. In these regular frames the particle masses are field-dependent and vanish in the infinite past, and the coefficient of the scalar kinetic term is negative without introducing an instability. This avoids the singularity theorems. By switching to the Einstein frame with constant particle masses and positive scalar kinetic term the singularity is introduced by the singular frame change. In the presence of higher curvature terms regular solutions with a scalar field have been found, for which cosmology in the Einstein frame first runs into a singularity and then moves away from a singularity. This clearly shows the artificial aspect of field singularities. 

If the universe can be extended beyond a singular point or not for particular cases \cite{Bars:2010zh,Bars:2013qna,Kamenshchik:2013dga,Kamenshchik:2016gcy} is under debate. For instance, one cannot remove anisotropies with a conformal transformation. Independently of the conformal frame some anisotropies will blow up in the past. If the singularity can be crossed \cite{Bars:2011aa,Kiefer:2018uyv,Kiefer:2019bxk}  one might argue that a singular change of variables (at the singular point) could cancel the singularity.

The black hole central singularity is more sophisticated than the FLRW one. It contains a component of shear reflected by the non-vanishing Weyl tensor squared. Since the Weyl tensor transforms multiplicatively under conformal transformations, such transformations are, in general, not enough to obtain a regular geometry. We will therefore consider disformal transformations of the metric that depend non-linearly on a scalar field $\phi$
\begin{align}\label{eq:disintro}
\bar g_{\mu\nu}=\omega^2(\phi) \bigl[ g_{\mu\nu}+D(\phi)\partial_\mu\phi\partial_\nu\phi \bigr]\,.
\end{align}
Depending on the solution for the scalar field such a transformation can indeed remove a black-hole-type central singularity from the geometry. For instance, in a similar direction Ref.~\cite{Naruko:2019gsi} showed that disformal transformations involving vector fields may be able to resolve spacetime singularities in Bianchi type-I anisotropic universes. Interestingly, their analysis could also be extended to Reissner-Nordstr\"{o}m black holes. Here, we will study black hole solutions with scalar hair.

A Schwarzschild black hole space-time can be written in the Kerr-Schild \cite{Visser:2007fj}, the Gordon \cite{Liberati:2018osj}, in the river forms \cite{Hamilton:2004au} and we will present a new way to write the Schwarzschild metric at the end of section \ref{subsec:horndenskiblackholes}. The first two are respectively written in Minkowski spherical coordinates as
\begin{align}\label{eq:kerrschild}
g^{\rm sch}_{\mu\nu}=\eta_{\mu\nu}+\frac{2m}{r}\partial_\mu\phi\partial_\nu\phi \,,\qquad \phi=-t+r\,,
\end{align}
and
\begin{align}\label{eq:gordon}
g^{\rm sch}_{\mu\nu}=\eta_{\mu\nu}+{\frac{1+2/r}{1+1/m}}\partial_\mu\varphi\partial_\nu\varphi \,,\qquad \varphi=-t+4\sqrt{1+r/2}\,,
\end{align}
where $\partial_\mu\phi\partial^\mu\phi=0$ is null-like and $\partial_\mu\varphi\partial^\mu\varphi<0$ is time-like.\footnote{Note that this would not be possible with a purely radial field as we would not be able to change the sign of the time component.} Generalizations to the Kerr metric involve a vector instead of the gradient of a scalar \cite{Visser:2007fj,Liberati:2018osj}. These forms are already suggestive for a removal of the singularity by a transformation of the type \eqref{eq:disintro}. Indeed, the metrics \eqref{eq:kerrschild} and \eqref{eq:gordon} look like a disformal transformation \cite{Bekenstein:1992pj} from Minkowski to Schwarzschild metrics, although the scalars $\varphi$ and $\phi$ are not dynamical yet--they are just mathematical objects.\footnote{It is interesting to note that disformal transformations map an exact solution of Einstein's equation to another \cite{BenAchour:2019fdf}.} From this simple example, it is clear that the Weyl tensor is in general not disformally invariant, \textit{e.g.} the Weyl tensor vanishes for Minkowski while it is not trivial for Schwarzschild. If we can promote these scalar fields to be solutions to Einstein and the field equations, \textit{e.g.} in a black hole with scalar hair, we would have a theory in which black holes are a matter of disformal frames. If so, it could also potentially mean that if we considered (the right) interaction between matter and the cosmological fields, matter fields might not see any singularity or that instead of a singularity the backreaction of the fields yields lumps of matter rather than black holes if the interaction is strong enough. See \textit{e.g.} Ref.~\cite{Bernardini:2009rc} -- although the detection of GWs from binary black holes \cite{Abbott:2016blz} tends to suggest otherwise.

Scalar hair is found in theories which present a spontaneous scalarization, \textit{e.g.} with a linear coupling to Gauss-Bonet term, or in shift-symmetric Horndeski theories \cite{Babichev:2013cya,Babichev:2017lmw,Babichev:2017guv,Saravani:2019xwx}. From the cosmological point of view, if there is a time varying scalar field on cosmological scales one expects that this field develops a profile if a black hole is present \cite{Gregory:2017sor,Gregory:2018ghc,Wong:2019yoc}. In the recent interest to analyze black hole solutions with scalar hair, we will study disformal transformations for static spherically symmetric solutions with a non-trivial scalar field. We will show that it is possible to alter significantly the geometry by a disformal transformation, \textit{e.g.} going to a frame with vanishing Weyl squared, and we will discuss whether the new frame can be regarded as a regular frame. To do so, we will consider models and solutions with a regular action to begin with and we will allow the transformation to be singular only at a singular point which might cancel the singularity. It should be noted that if the metric transformation is regular then the singularity of the solution cannot be removed.

Disformal transformations have been widely studied in cosmology and are motivated from a higher dimensional perspective in brane-cosmology, \textit{e.g.} a moving brane \cite{KOIVISTO:2013jwa,Koivisto:2013fta} or an effective description of the fact that gravitons could travel in one higher dimension \cite{Ishihara:2000nf}. Interestingly, they have been shown to leave the form of the Horndeski lagrangian \cite{Horndeski1974} (and beyond \cite{Gleyzes:2014dya,Langlois:2015cwa,Crisostomi:2016czh,Langlois:2017mxy}) invariant \cite{Bettoni,Zumalacarregui:2013pma} and that, in cosmology, physical predictions are unchanged \cite{Deruelle:2014zza,Domenech:2015tca}. Also see how disformal transformations describe the effective causal structure for the scalar field in K-essence \cite{Babichev:2007dw} and hairy black holes \cite{Benkel:2018qmh}. It has been argued \cite{Wetterich:2014bma} that disformal transformations can be used to transform a very wide class of higher derivative theories with a physical scalar degree to standard scalar-tensor as Brans-Dicke type theories. Recently, disformal trasnformations have been used to generate new solutions in DHOST theories \cite{BenAchour:2019fdf}. Therefore, disformal transformations play an important role in higher derivative theories. In this paper, we study the applications of disformal transformations to static spherically symmetric systems.

The paper is organized as follows. In Sec.~\ref{sec:hawkingturok} we review the Hawking-Turok instanton which motivates the study of finding a regular frame for black holes. In Sec.~\ref{sec:sphericallysymmetric}, we study disformal transformations in static spherically symmetric metrics with scalar hair. First we draw some general conclusions and then focus on three particular examples within Horndeski theory. In particular, we discuss a critical cosmon lump which is a singular black hole type solution in the Einstein frame, while being a regular solution in the scaling frame. We conclude our work in Sec.~\ref{sec:conclusions}. Details on the calculations and a disformal invariant tensor can be found in the appendices. We work in (reduced) Planck units where $M_{pl}=1/8\pi G=1$ and $\hbar=c=1$.

\section{Warm up: cosmological example \label{sec:hawkingturok}}

In this section, we review the Hawking-Turok instanton \cite{Hawking:1998bn}, where an open universe is created by analytical continuation of a singular Euclidean manifold. As we will see, this example is relevant due to its possible embedding in a regular 5D manifold \cite{Garriga:1998ri}. Hawking and Turok considered a canonical scalar field $\phi$ in a potential $V(\phi)$ whose action in the Euclidean signature reads
\begin{align}
S_E=\int d^4 x \sqrt{g}\left[-\frac{1}{2}{\cal R}+\frac{1}{2}\nabla_\mu\phi\nabla^\mu\phi+V(\phi)\right]\,,
\end{align}
where $R$ is the Ricci scalar and $g$ is the metric determinant. There is an $O(4)$ symmetric solution in which the metric is given by
\begin{align}
ds^2=d\sigma^2+b^2(\sigma)\left(d\chi^2+\sin^2\chi \,d\Omega^2\right) \,, \qquad \phi=\phi(\sigma)\,,
\end{align}
where 
\begin{align}
b(\sigma)\approx
			\begin{cases}
               \sigma \qquad &(\sigma\sim 0)\\
               \left(\sigma_f-\sigma\right)^{1/3} \qquad &(\sigma\sim \sigma_f)
            \end{cases} \,,
\qquad
\phi\approx
			\begin{cases}
               \sigma^2/2 \qquad &(\sigma\sim 0)\\
              - \sqrt{(2/3)} \, \ln\left(\sigma_f-\sigma\right)\qquad &(\sigma\sim \sigma_f)
            \end{cases}\,.
\end{align}
The open universe is obtained after a Wick rotation of the variable $\chi$. Although the Ricci scalar near $\sigma\sim \sigma_f$ blows up like ${\cal R}\sim (\sigma_f-\sigma)^{-2}$, the on-shell action is regular if $V$ does not grow faster than $b^3$ as $\sigma\to\sigma_f$. To be clearer, we can evaluate the action using the trace of Einstein's equations:
\begin{align}
 {\cal R}_{\mu \nu} - \frac{1}{2} {\cal R} \, g_{\mu \nu} = \pa_\mu \phi \pa_\nu \phi - \left[ \frac{1}{2} \nabla_\mu \phi \nabla^\mu \phi + V \right] g_{\mu \nu} \quad \to \quad
 - {\cal R} = - \nabla_\mu \phi \nabla^\mu \phi - 4 V 
 \,,
\end{align}
 and we find that
\begin{align}
S_E=-\int d^4x \, b^3 \, V(\phi)\,.
\end{align}
However, as it is usually the case in cosmology, the singularity in the Ricci scalar is a conformal singularity since the metric can be written as
\begin{align}
ds^2=b^2(\overline \sigma)\left(d\overline\sigma^2+d\chi^2+\sin^2\chi \,d\Omega^2\right)\equiv b^2(\overline \sigma) \, d\overline s^2 \,,
\end{align}
where $d\overline\sigma\equiv b^{-1}d\sigma$ and we have introduced a new metric $\overline g_{\mu\nu}\equiv b^{-2} g_{\mu\nu}$. If we study this transformation near the singularity we find that $\overline\sigma=(3/2) \left(\sigma_f-\sigma\right)^{2/3}$ such that a transformation given by
\begin{align}
b(\phi)={\rm e}^{\phi/\sqrt{6}}\,,
\end{align}
is enough to obtain a frame where there is no singularity in the geometry as $\overline \sigma\to 0$. It should be noted that the transformation is only singular at the singularity, that is at $\sigma\sim\sigma_f$ ($\overline \sigma\sim 0$). The action in terms of the barred metric reads
\begin{align}\label{eq:likegarriga}
S_E=\int d^4 x \sqrt{\overline g}\left[-\frac{1}{2}{\mathrm{e}}^{-\sqrt{(2/3)} \, \phi}\overline {\cal R}+\overline V(\phi)\right]\,,
\end{align}
where $\overline {\cal R}$ is the Ricci scalar corresponding to the metric $\overline g_{\mu\nu}$ and 
\begin{align}
\overline V(\phi)\equiv{\mathrm{e}}^{-2\sqrt{(2/3)} \, \phi} \, V(\phi) \,.
\end{align}
Therefore, we have removed the singularity from the geometry and we possibly left it in the scalar field. The regularity of the action is now recasted as a condition on the regularity of the potential $\overline V$. So far this is common from conformally flat metrics, what is really interesting is that this action corresponds to the dimensional reduction of the 5D regular euclidean instanton \cite{Garriga:1998ri}. In other words, from the higher dimensional point of view the instanton is not singular, rather the space ends at that point.

\textcolor{black}{We have found a frame where the action is regular but the field is still singular.} Now the question is then: can we make the field regular as well by a field redefinition? Since the action lacks of the standard kinetic term for the scalar field it is not possible to argue this at present. Nevertheless, we can do another conformal transformation that does not spoil the singularity-free geometry and that gives a kinetic term to the field. For that purpose, we can use a conformal transformation which goes to unity at the singularity, \textit{e.g.}
\begin{align}
\overline g_{\mu\nu}=\Omega^2 \widetilde g_{\mu\nu}=\left(1+\beta {\rm e}^{-\alpha\sqrt{2/3} \, \phi}\right)^2\widetilde g_{\mu\nu}\,,
\end{align}
and then the action becomes
\begin{align}
S_E=\int d^4 x \sqrt{\widetilde g}\left[-\frac{1}{2}{\mathrm{e}}^{-\sqrt{2/3} \, \phi}\Omega^2 \widetilde {\cal R} +3{\mathrm{e}}^{-\sqrt{2/3} \, \phi}\Omega^2\widetilde\nabla_\mu\phi\widetilde\nabla^\mu\phi\,\frac{\partial\ln\Omega}{\partial\phi}\left(\sqrt{\frac{2}{3}}-\frac{\partial\ln\Omega}{\partial\phi}\right)+\Omega^4\overline V(\phi)\right]\,.
\end{align}
One can check that the Ricci scalar near the singular point $\overline \sigma\sim 0$ goes like $\widetilde {\cal R}\propto \overline\sigma^{\alpha-2}$ and so it is regular for $\alpha>2$. Note that we may not worry about the prefactors since $\Omega\to1$ towards the singular point. We can see that now the canonically normalized field is regular as well and it goes as
\begin{align}
d\varphi^2\equiv 6 \,d\phi^2\,{\mathrm{e}}^{-\sqrt{2/3} \, \phi}\Omega^2\frac{\partial\ln\Omega}{\partial\phi}\left(\sqrt{\frac{2}{3}}-\frac{\partial\ln\Omega}{\partial\phi}\right) \,.
\end{align}
Near the singular point thus reads
\begin{align}
d\varphi\sim \,d\phi\,{\rm e}^{- (1+\alpha) \phi/\sqrt{6}} \quad \to \quad \varphi\sim {\rm e}^{- (1+\alpha) \phi/\sqrt{6}} \sim \overline\sigma^{ (1+\alpha)/2} \,,
\end{align}
{where we chose for simplicity that $\beta=-1/4\alpha$} with $\varphi\to 0$ as $\phi\to\infty$, the kinetic term $\partial_{\overline\sigma}\varphi\to 0$, is regular.
{With these new variables, the Lagrangian near the singular point reads
\begin{align}
 {\cal L} \sim - \frac{1}{2} \varphi^{2/(1 + \alpha)} \widetilde {\cal R}
 + \frac{1}{2} \widetilde\nabla_\mu\varphi\widetilde\nabla^\mu\varphi+\overline V(\varphi) \,.
\end{align}}
Thus we have found a frame where the geometry, the scalar field, the kinetic and potential terms are all regular.

\section{Static spherically symmetric metric \label{sec:sphericallysymmetric}}

We have seen that in cosmological set-ups where the metric is conformally flat, one can rewrite the singularity in the geometry as a singularity in the scalar field or, perhaps, one may find a regular frame. It is then worth to study whether this is restricted to conformally flat space-times only. For that purpose, we consider a static spherically symmetric metric given by
\begin{align}\label{eq:sphericallysymmetricmetric}
ds^2=-h(r)dt^2+f(r)dr^2+r^2d\Omega^2 \,,
\end{align}
where $f,h$ are arbitrary functions of $r$ and $fh>0$ everywhere. The squared Weyl tensor for this metric is given by
\begin{align}\label{eq:weyl}
{\cal C}_{\alpha\beta\sigma\rho}{\cal C}^{\alpha\beta\sigma\rho}=\frac{1}{12f^2r^4}\left[\frac{r^2h'^2}{h^2}+4f-4-\frac{2rf'}{f}+\frac{rh'}{h}\frac{rf'}{f}+\frac{2rh'}{h}-2\frac{r^2h''}{h}\right]^2\,.
\end{align}
For known black hole solutions this invariant diverges for $r\to0$. We will investigate if a singularity of this type in the geometry can be removed by a field transformation. For the moment, we leave the exact form of the action unspecified but we assume that the model has a static spherically symmetric solution with scalar hair $\phi=\phi(t,r)$. We investigate here if a different choice of metric can remove the singularity from the geometry. 

With respect to a conformal transformation the Weyl tensor transforms multiplicatively,
\begin{align}\label{eq:multiplicatively}
\bar g_{\mu\nu}=\omega^2(\phi)g_{\mu\nu}\,,\qquad \bar {\cal C}_{\alpha\beta\sigma\rho} =\omega^2(\phi)\,{\cal C}_{\alpha\beta\sigma\rho}\,,\qquad \bar {\cal C}_{\alpha\beta\sigma\rho}\bar{\cal C}^{\alpha\beta\sigma\rho}=\omega^{-4}(\phi)\,{\cal C}_{\alpha\beta\sigma\rho}{\cal C}^{\alpha\beta\sigma\rho} \,,
\end{align}
and the Ricci scalar transforms as
\begin{align}
\bar{\cal R}=\omega^{-2}(\phi)\left({\cal R}-6\Box\ln \omega-6\nabla^\mu\ln \omega\nabla_\mu\ln \omega\right)\,.
\end{align}
The combination $g^{1/2}{\cal C}_{\alpha\beta\sigma\rho}{\cal C}^{\alpha\beta\sigma\rho}$ is invariant under conformal transformations. If this combination shows a singularity that cannot be absorbed by a general coordinate transformation, this singularity cannot be removed by any conformal transformation. One might do so, however, with a generalized metric transformation called disformal transformation, which reads
\begin{align}\label{eq:disformaltransformation}
\bar g_{\mu\nu}=\omega^2(\phi)\bigl[g_{\mu\nu}+D(\phi)\partial_\mu\phi\partial_\nu\phi\bigr]\,,
\end{align}
where $\omega$ and $D$ are functions of $\phi$ and its derivatives. The functions $\omega(\phi)$ and $D(\phi)$ are chosen such that the spherically symmetric solutions only yield a radial dependence for $\omega$ and $D$. With $\omega=1$ we are modifying the direction along the gradient of the scalar field only. 

In order to proceed, we will assume that the solution for the scalar field satisfies the following ansatz,
\begin{align}\label{eq:notation}
\phi(t,r)=q \, t+\psi(r)\,,
\end{align}
which is often the case for static spherically symmetric solutions with scalar hair \cite{Babichev:2013cya}. The metric $\bar g_{\mu\nu}$ in the new frame is computed from $g_{\mu\nu}$ by inserting in Eq.~\eqref{eq:disformaltransformation} the scalar field solution Eq.~\eqref{eq:notation}. After the disformal transformation \eqref{eq:disformaltransformation}, the barred metric can also be written as a spherically symmetric metric given by
\begin{align}\label{eq:sphericallysymmetricmetric2}
d\bar s^2=-\bar h(\bar r)d\bar t^2+\bar f(\bar r)d\bar r^2+\bar r^2d\Omega^2 \,,
\end{align}
where
\begin{align}\label{eq:transformationrules}
\bar r^2=\omega^2 r^2 \,,\qquad \bar h=\omega^2\left(h-Dq^2\right) \,,\qquad \bar f=\left(f+\frac{Dh\psi'^2}{h-Dq^2}\right)\left(1+\frac{\partial \ln \omega}{\partial \ln r}\right)^{-2} \,,
\end{align}
with $\psi'\equiv \partial_r\psi$. The disformal transformation is accompanied by a coordinate transformation, with
\begin{align}
d\bar t=dt-\frac{Dq\psi'}{h-Dq^2} dr\,.
\end{align}
These are the main formulas that we will use in the examples below. Furthermore, we will require that the metric transformation is regular and invertible, except at the singular point. Thus, we will require that $\omega>0$, $1+D(\partial\phi)^2>0$ except at any singularity. For future use, it should be noted that the latter requirement is equivalent to the condition that $\bar h \bar f>0$.

We can study in general whether it is possible at all to transform to a Minkowksi barred metric or a Schwarzchild barred metric. In the former case, we might be able to choose $\omega$ and $D$ such that the resulting metric is Minkowski space-time if $\bar h = \bar f = 1$. We can draw two general conclusions. First, if there is a singularity inside a horizon in the metric $g_{\mu\nu}$, we see from Eq.~\eqref{eq:transformationrules} that in order to go to a Minkowski frame one needs $q\neq 0$. For $q=0$ the fact that $h<0$ inside the horizon implies $\omega^2<0$, which is not physically sound. Thus, to remove singularities inside a horizon with a scalar field one needs a time-dependent scalar field with hair. Second, a naked singularity can be removed by a scalar field with radial profile only if $h>0$ everywhere. We will present below example where these conditions are met. We will find that indeed a disformal transformation \eqref{eq:disformaltransformation} can remove the singularity from the geometry.

On the other hand, we may ask which type of metric can be trasnformated by a disformal transformation to a singular Schwarzschild metric. We may then start with a regular metric and induce the singularity only by the transformation. We consider here a static setting, $q=0$. A Schwarzschild barred metric may be possible if there is a solution to $\bar h=1-\mu/\bar r=\omega^2 h$ and $\bar r =|\omega| r$. In other words, there must be a solution to the algebraic equation 
\begin{align}\label{eq:horizon}
h=\frac{1}{\omega^2}\left(1-\frac{\mu}{\omega r}\right)\quad\Rightarrow \quad hr^{-2}=\bar h \bar r^{-2}\,,
\end{align}
where we assumed in the last step that $w>0$. Now, if the original metric has a singularity at $r=0$ the condition that there is a solution to $\bar h=0$ when $r\to 0$ is that $h\,r^{-2}\to0$. Furthermore, since by assumption $\bar r\to\mu$ when $r\to 0$ we need that $\omega\propto r^{-1}$. In that case, we can go to a Schwarzschild barred metric. The same conclusion can be drawn by expanding the transformation around $r\sim 0$. 

Introducing a singularity through a field transformation is easy. For example, consider that the metric \eqref{eq:sphericallysymmetricmetric} is perfectly regular, that is when $r\to 0$ we have $h,f\to {\rm constant}$ which could be the interior of a star. Now, if we take a singular transformation Eq.~\eqref{eq:disformaltransformation} with $\omega^2=1/r$ and $D=0$, we see using Eq.~\eqref{eq:transformationrules} that as $r\to 0$ one has $\bar r\to 0$, $\bar h\propto \omega^2\to\infty$ and $\bar f \to f/2={\rm constant}$. Thus, we have introduced a singularity from a singular metric transformation.

Regarding the ``physical'' nature of a singularity, it has been suggested in the past that a singularity is physical if the Weyl tensor is singular. Since the Weyl tensor transforms multiplicatively under conformal transformations \eqref{eq:multiplicatively}, singularities in the Weyl tensor can often not be removed by a conformal transformation. In other words, the Weyl tensor would blow up if the shear of the metric Eq.~\eqref{eq:sphericallysymmetricmetric} blows up. The Weyl tensor is not disformal invariant and we will show by examples that the singularity can indeed be transformed away by a disformal transformation.

The change of the Weyl tensor can be easily seen from the transformation of the Riemann tensor. In a simple exercise, one finds for the transformation \eqref{eq:disformaltransformation} with $\omega=D=1$ that \cite{Zumalacarregui:2013pma}
\begin{align}
\bar {\cal R}_{\alpha\beta\sigma\rho}={\cal R}_{\alpha\beta\sigma\rho}+\frac{2}{1+\partial_\mu\phi\partial^\mu\phi}\nabla_{\sigma}\nabla_{[\alpha}\phi\nabla_{\beta]}\nabla_{\rho}\phi \,,
\end{align}
where the $[\,,\,]$ is the normalized anti-symmetrization of indices. Since 
\begin{align}
{\cal C}_{\alpha\beta\sigma\rho}={\cal R}_{\alpha\beta\sigma\rho}+{\cal R}_{\beta[\sigma}g_{\rho]\alpha}-{\cal R}_{\alpha[\sigma}g_{\rho]\beta}+\frac{1}{3}{\cal R}g_{\alpha[\sigma}g_{\beta]\rho} \,,
\end{align}
there will not be any term with second derivatives of $\phi$ in $\bar {\cal R}_{\alpha\beta}$ or $\bar {\cal R}$ capable of canceling the one coming from $\bar {\cal R}_{\alpha\beta\sigma\rho}$, except for very particular cases.\footnote{For example, in the cases where the second derivative is related to first derivatives of scalar fields by $\nabla_\mu\nabla_\nu\phi= A_1\nabla_\mu\phi\nabla_\nu\phi+2A_2\nabla_{(\mu}\varphi\nabla_{\nu)}\phi+A_3\nabla_\mu\varphi\nabla_\nu\varphi$ with $\nabla_\mu\varphi\nabla^\mu\phi=0$. This is for example the case for a geodesic scalar field.} \textcolor{black}{That is to say that the second derivatives of $\phi$ contained in $\bar {\cal R}_{\alpha\beta}=\bar g^{\sigma\rho}\bar {\cal R}_{\sigma\alpha\rho\beta}$ and $\bar {\cal R}=\bar g^{\alpha\beta}\bar {\cal R}_{\alpha\beta}$ will be contracted either with the metric $g_{\mu\nu}$ or with the first derivatives of the field and will not have any free index to cancel that of $\bar {\cal R}_{\alpha\beta\sigma\rho}$. For example $\bar {\cal R}\supset \left(\Box\phi\right)^2-\nabla_\mu\nabla_\nu\phi\nabla^\mu\nabla^\nu\phi$.} Thus, the Weyl tensor is in general not invariant under a disformal transformation. Then the question of whether the geometry can be made regular is a question of whether a disformal transformation is able to make the Weyl tensor regular. For the transformed metric in the barred frame one must replace all quantities in \eqref{eq:weyl} by their barred counterparts. In general, ${\cal C}_{\alpha\beta\sigma\rho}{\cal C}^{\alpha\beta\sigma\rho}$ differs from $\bar {\cal C}_{\alpha\beta\sigma\rho}\bar {\cal C}^{\alpha\beta\sigma\rho}$ and the squared Weyl tensor is not invariant. However, one may build a tensor using the scalar field involved in the transformation which is disformal invariant (see Appendix \ref{app:disformalinvariant}).

\subsection{Cosmon lumps \label{subsec:cosmonlumps}}

We will first study a spherically symmetric solution of a massless scalar field \cite{Buchdahl:1959nk,Wyman:1981bd}, which has been dubbed cosmon lump \cite{Wetterich:2001cn}. This solution has been shown to be unstable in Ref.~\cite{Krueger:2008nq} but nevertheless is the simplest example of a solution with scalar hair. Furthermore, it has to be seen if the inclusion of matter with a disformal coupling might stabilize the solution. The action is given by Einstein gravity coupled to a free massless scalar field,
\begin{align}\label{eq:cosmonaction}
S=\int d^4x \sqrt{-g} \,\frac{1}{2} \left( {\cal R}- X\right) \,,\quad X=\nabla_\mu\phi\nabla^\mu\phi\,.
\end{align}
The action vanishes on-shell after using the trace of Einstein's equations, ${\cal R} = X$ and hence the action is regular. The spherically symmetric solution is given by
\begin{align}\label{eq:hfphicosmon}
h(r)=\left(\frac{\rho-\rho_H}{\rho+\rho_H+R_s}\right)^{\delta} \,, \qquad f(r)=\rho^{-2}\left(\rho-\rho_H\right)\left(\rho+\rho_H+R_s\right) \,,\qquad \phi=\phi_0+\frac{\gamma}{\sqrt{2}}\ln h(r)\,,
\end{align}
with
\begin{align}\label{eq:rrhocosmon}
r=\left(\rho-\rho_H\right)^{(1-\delta)/2}\left(\rho+\rho_H+R_s\right)^{(1+\delta)/2} \,.
\end{align}
We have introduced
\begin{align}
\delta\equiv\left(1+\gamma^2\right)^{-1/2} \,,\qquad \rho_H\equiv\frac{R_s}{2\delta}\left(1-\delta\right) \,, \qquad \delta<1\,,
\end{align}
where $\phi_0$, $R_s$ and $\gamma$ are integration constants. The Schwarzschild solution is approached for $\gamma\to0$, $\delta\to1$ and $R_s$ corresponds to the usual Schwarzschild radius $R_s=2GM$. Useful relations to derive this solution are given by
\begin{align}\label{eq:useful}
\frac{\partial\ln \rho}{\partial\ln r}=h \qquad{\rm and}\qquad \frac{\partial\ln h}{\partial\ln r}=\frac{R_s}{\rho}\,.
\end{align}
From now on we will set $\phi_0=0$ for simplicity unless stated otherwise.

This kind of solution is also known as singular Brans-Dicke type I solutions \cite{Faraoni:2016ozb} (see Appendix \ref{app:cosmonlumps}). For large radii, this solution yields approximately the Schwarzschild metric but there is a naked singularity at $r=0$ which corresponds to $\rho=\rho_H$. The main difference to the black hole solution is the smoothening near the Schwarzschild radius--there is no longer a horizon. For many observational aspects the cosmon lump is difficult to distinguish from a black hole \cite{Wetterich:2001cn,Krueger:2008nq}. The limit $\gamma\to0$ ($\delta\to1$) corresponds to the Schwarzschild solution, but the limit is not continuous since for $\gamma=0$ this solution only covers the patch of the isotropic coordinates (see Appendix \ref{app:cosmonlumps}). The squared Weyl tensor for this solution is given by
\begin{align}
{\cal C}_{\alpha\beta\sigma\rho}{\cal C}^{\alpha\beta\sigma\rho}=
\frac{1}{3}\left[\frac{R_s\left(6\rho-R_s\gamma^2\right)}{r^2f\rho^2}\right]^2=\frac{1}{3}\left[\frac{2R_s\left(3\rho-[1+\delta^{-1}]\rho_H\right)}{(\rho-\rho_H)^{2-\delta}(\rho+\rho_H+R_s)^{2+\delta}}\right]^2 \,.
\end{align}
It is divergent for $r\to0$ (or $\rho\to\rho_H$) as $(\rho-\rho_H)^{-2(2-\delta)}$. At the singularity we have $\rho=\rho_H\neq0$. An exceptional case is $\delta=1/2$ for which the squared Weyl tensor blows up as $(\rho-\rho_H)^{-1}$ because for $\delta=1/2$ ($\gamma^2=3$) the numerator vanishes as $(\rho-\rho_H)^{2}$ since $R_s=2\rho_H$. Multiplying with $\sqrt{-g}$ one finds the conformally invariant expression (in four dimensions)
\begin{align}
\sqrt{-g}{\cal C}_{\alpha\beta\sigma\rho}{\cal C}^{\alpha\beta\sigma\rho}=
\frac{r^3h\sin\theta}{3\rho}\left[\frac{R_s\left(R_s\gamma^2-6\rho\right)}{r^2f\rho^2}\right]^2\,.
\end{align}

This solution can be transformed to the regular geometry of Minkowski space by the choice of a different metric
\begin{align}\label{eq:distranscosmon}
\bar g_{\mu\nu}=\omega^2(\phi)\bigl[ g_{\mu\nu}+D(\phi)\partial_\mu\phi\partial_\nu\phi \bigr] \,,
\end{align}
where $\omega(\phi)$ and $D(\phi)$ are respectively given by
\begin{align}
\omega(\phi)={e}^{-\frac{\phi}{\sqrt{2}\gamma}} \,, \qquad D(\phi)=\frac{R_s^2}{2\gamma^2\delta^4}\frac{e^{\sqrt{2}\frac{1-\delta}{\gamma\delta}\phi}}{1-e^{\sqrt{2}\frac{\phi}{\gamma\delta}}}\left[\left(1-2\delta\right)^2+e^{\sqrt{2}\frac{\phi}{\gamma\delta}}\left(1+2\delta\right)^2\right]\,.
\end{align}
We will see shortly that this field dependent transformation reduces to the conditions \eqref{eq:transformationrules}.
We can use the solution $\phi(\rho)$ in Eq.~\eqref{eq:hfphicosmon} with $\phi_0=0$ to find the radial dependence of $\omega(\rho)$ and $D(\rho)$ for a given solution with the same $\gamma$, $R_s$. We obtain
\begin{align}
\omega(\rho)={e}^{-\frac{\phi(\rho)}{\sqrt{2}\gamma}}=\left(\frac{\rho-\rho_H}{\rho+\rho_H+R_s}\right)^{-\delta/2} \,,
\end{align}
and
\begin{align}
D(\rho) &= \frac{R_s^2}{2\gamma^2\delta^4}\frac{e^{\sqrt{2}\frac{1-\delta}{\gamma\delta}\phi}}{1-e^{\sqrt{2}\frac{\phi}{\gamma\delta}}}\left[\left(1-2\delta\right)^2+e^{\sqrt{2}\frac{\phi}{\gamma\delta}}\left(1+2\delta\right)^2\right]\nonumber\\&=\frac{1}{2\delta^2\gamma^2}\left[1-8\delta^2\frac{\rho}{R_s}\right]\left(\rho-\rho_H\right)^{1-\delta}\left(\rho+\rho_H+R_s\right)^{1+\delta}\,.
\end{align}
These results actually correspond to Eq.~\eqref{eq:transformationrules} with  $\bar h=\bar f=1$ and $q=0$, explicitly given by
\begin{align}
\omega^2(r)=h^{-1}\quad{\rm and }\quad D(r)=\left[\left(1-\frac{1}{2}\frac{\partial\ln h}{\partial \ln r}\right)^{2}-f\right]\left(\frac{\partial\phi}{\partial r}\right)^{-2}\,.
\end{align}
We also have that $\bar r = h^{-1/2}r$. We have therefore demonstrated that for every singular cosmon lump with given parameters $\gamma$ and $R_s$ there exists a choice of the metric $\bar g_{\mu\nu}$ \eqref{eq:distranscosmon} which describes a regular flat space geometry. As it should be, this transformation is singular only at $r=0$ where 
\begin{align}
\omega=h^{-1/2}\sim \left(\frac{r}{\rho_H}\right)^{- \delta/(1-\delta)}\to\infty\,.
\end{align}
The disformal factor $D$ vanishes for $r\to0$ as $\phi'\propto 1/r$ and $f\sim \left({r}/{\rho_H}\right)^{2/(1-\delta)}\to 0$ for all values of $\delta$, since $\delta<1$.
We are not ``postponing'' the singularity since $\bar r \sim r^2/\rho_H\to 0$. Since the transformed metric is flat there is obviously no geometrical singularity at $\bar r =0$. If we keep the original definition of the scalar field, it is still singular at $\bar r=0$. However, we can introduce a new field which is regular at $\bar r=0$, \textit{e.g.}, $\varphi\equiv e^{\phi/\gamma}=h$. Independently of the choice of the field, we know that the energy momentum tensor of the field will be finite by the modified Einstein's equations.

It should be noted that the disformal transformation \eqref{eq:distranscosmon} only renders the metric \eqref{eq:hfphicosmon} regular for a given solution, \textit{i.e.} given a fixed $R_s$ and $\delta$. In other words, if we applied the transformation to Minkowski spacetime for a particular solution, say $R_s=R_{s,1}$ and $\delta=\delta_1$, to the metric of another solution with $R_s=R_{s,2}$ and $\delta=\delta_2$, we would not be able to obtain a regular metric. Looking only at the conformal factor, we conclude that we will not be removing the original singularity but we may introduce a new one.

Before ending this subsection, we may consider in a second example if it is possible to switch the naked singularity for a black hole horizon. If we try to obtain a barred metric which is Schwarzschild by choosing 
\begin{align}
\bar h= \frac{1}{\bar f} =1-\frac{\mu}{\bar r} \,,
\end{align}
in Eq.~\eqref{eq:transformationrules} we find that Eq.~\eqref{eq:horizon} has a solution only when $1>\delta>1/2$ which is given by $\omega\sim\mu/r$ near $r\sim 0$ since in that case we have that $\omega^2h\sim h\,r^{-2}\propto r^{2\frac{2\delta-1}{1-\delta}}\to 0$. If $\delta<1/2$ then $h\,r^{-2}\to\infty$ and there is no solution to Eq.~\eqref{eq:horizon} and no frame with a Schwarzschild metric. Therefore, for the cosmon lump solution it is possible to introduce a Schwarzschild like horizon when $1>\delta>1/2$ which corresponds to the naked singularity of the cosmon solution at $r=0$.

We will not discuss here the specific form of the resulting action after the disformal transformation \eqref{eq:disformaltransformation} for two reasons, although we have derived the general transformation rules in App.~\ref{app:disformulas} and presented the explicit expression in  App.~\ref{eq:cosmondisfaction}. First and most important, the resulting action falls within Horndeski theory, where there is no clear notion of a canonical scalar field. The canonical definition is usually relevant as its first derivatives are related to the energy density of the field. However, due to the presence of higher derivatives in the Horndeski action it is not clear how to define the energy momentum tensor of the scalar field as it mixes with gravity. Thus, even though the action vanishes on-shell, it is difficult to argue that an arbitrary field redefinition which makes the field and its derivatives regular, renders each term in the action separately regular. 
Second, the involved form of the action does not yield any significant physical insight. Nevertheless, there is a particular case where we can study the resulting frame exactly, giving further support to the regularity of the geometry and the field content.

\subsection{Critical cosmon lumps\label{subsec:critical}}
We have already seen that for the particular case
\begin{align}
\delta=1/2 \,,\qquad \gamma^2=3 \,,\qquad\rho_H=R_s/2\,,
\end{align}
the behavior of the squared Weyl tensor at the singularity is exceptional. We call the solution with $\delta=1/2$ the critical cosmon lump and discuss a few of its properties. With
\begin{align}
r^2=\left(\rho-\rho_H\right)^{1/2}\left(\rho+3\rho_H\right)^{3/2}\,,\qquad h=\left(\rho-\rho_H\right)^{1/2}\left(\rho+3\rho_H\right)^{-1/2}\,,\qquad \rho^2f=\left(\rho-\rho_H\right)\left(\rho+3\rho_H\right)
\end{align}
one finds for the metric 
\begin{align}\label{eq:criticalcosmonmetric}
ds^2=\left(\frac{\rho-\rho_H}{\rho+3\rho_H}\right)^{1/2}\left[-dt^2+\frac{\rho+3\rho_H}{\rho-\rho_H}d\rho^2+\left({\rho+3\rho_H}\right)^2d\Omega^2\right]\,,
\end{align}
where we used the relation \eqref{eq:useful} to express everything in terms of $\rho$. For this metric, the Ricci scalar and the Weyl tensor squared respectively read
\begin{align}
{\cal R}=\frac{6\rho_H^2}{\left({\rho-\rho_H}\right)^{3/2}\left({\rho+3\rho_H}\right)^{5/2}} \,,\qquad {\cal C}_{\alpha\beta\sigma\rho}{\cal C}^{\alpha\beta\sigma\rho}=\frac{48\rho_H^2}{\left({\rho-\rho_H}\right)\left({\rho+3\rho_H}\right)^5}\,.
\end{align}
As expected, both the Ricci scalar and the Weyl tensor squared diverge at the singularity $\rho=\rho_H$, that is
\begin{align}
{\cal R}\to \frac{96\rho_H^4}{r^6}\propto\left(\rho-\rho_H\right)^{-3/2}\qquad{\rm and}\qquad {\cal C}_{\alpha\beta\sigma\rho}{\cal C}^{\alpha\beta\sigma\rho}\to\frac{3}{r^4}\propto\left(\rho-\rho_H\right)^{-1} \,.
\end{align}
The conformally invariant combination,
\begin{align}
d^4x\sqrt{-g}\,{\cal C}_{\alpha\beta\sigma\rho}{\cal C}^{\alpha\beta\sigma\rho}
=\frac{48\rho_H^2}{\left({\rho+3\rho_H}\right)^6}\left(\frac{\rho+3\rho_H}{\rho-\rho_H}\right)^{1/2}(\rho+3\rho_H)^2 dt \,d\rho \,d\Omega\,,
\end{align}
is free of singularities since the singular term can be absorbed with a redefinition of the radial coordinate $\rho$.

The exceptional property of this solution is that the Weyl tensor squared diverges as $(\rho-\rho_H)^{-1}$, which is exactly the inverse square of the conformal factor in the metric \eqref{eq:criticalcosmonmetric}. This suggests that a conformal transformation (or the disformal transformation \eqref{eq:disformaltransformation} with $D=0$), which rescales the Weyl tensor squared, might be able to remove the singularity at least in the Weyl tensor squared. To be more precise, we rewrite the metric as
\begin{align}\label{eq:metricregular}
ds^2= \omega^{-2}d\bar s^2\,, \qquad \omega={\rm e}^{-\frac{1}{\sqrt{6}}\phi}\,.
\end{align}
Insertion of the cosmon lump solution yields
\begin{align}
\omega=\left(\frac{\rho-\rho_H}{\rho+3\rho_H}\right)^{-1/4}\,,
\end{align}
and the Weyl tensor squared is rescaled according to
\begin{align}
\bar{\cal C}_{\alpha\beta\sigma\rho}\bar{\cal C}^{\alpha\beta\sigma\rho}=\omega^{-4}{\cal C}_{\alpha\beta\sigma\rho}{\cal C}^{\alpha\beta\sigma\rho}=\frac{48\rho_H^2}{\left({\rho+3\rho_H}\right)^6}\,.
\end{align}
With the conformal transformation \eqref{eq:metricregular} we have obtained a metric with a regular Weyl tensor squared. Moreover, transforming the Ricci scalar as well, we find that $\bar {\cal R}=0$. The barred metric 
\begin{align}\label{eq:criticalcosmonmetric2}
d\bar s^2=-dt^2+\frac{\rho+3\rho_H}{\rho-\rho_H}d\rho^2+\left({\rho+3\rho_H}\right)^2d\Omega^2 \,,
\end{align}
has only a coordinate singularity at $\rho=\rho_H$.

The regularity of the barred metric can be seen by rewriting the metric in the isotropic coordinate $R$ (see App.\ref{app:cosmonlumps} for more details) related to $\rho$ by
\begin{align}
\rho=\rho_H+R\left({1-\frac{\rho_H}{R}}\right)^2\,.
\end{align}
The association between $R$ and $\rho$ is not unique since $R$ and $R'=\rho_H^2/R$ correspond to the same $\rho$. For $\rho=\rho_H$ one has $R=\rho_H$.
In this choice of coordinates the barred metric reads
\begin{align}\label{eq:metriciso}
d\bar s^2=-dt^2+\left({1+\frac{\rho_H}{R}}\right)^4\left(dR^2+R^2d\Omega^2\right)\,.
\end{align}
This metric is indeed regular at $R=\rho_H$ and it can be extended to $R<\rho_H$. In fact, the barred metric \eqref{eq:metriciso} has the symmetry $R\to R'={\rho_H^2}/{R}$ and, therefore, the region $0<R\leq\rho_H$ is a copy of the patch $\rho_H\leq R< \infty$. Both describe the same $\rho$. Comparing with Ref.~\cite{Faraoni:2016ozb} the barred metric \eqref{eq:metriciso} corresponds to a particular case of Brans Dicke type I solutions with $w=6(C+1)/C^2$, $C=\lambda+2$ and then take the limit $\lambda\to\infty$. The regularity of the geometry is also clear from the Weyl tensor squared for the barred metric which is given by
\begin{align}
\bar {\cal C}_{\alpha\beta\sigma\rho}\bar {\cal C}^{\alpha\beta\sigma\rho}=\frac{48\rho_H^2}{R^6}\left({1+\frac{\rho_H}{R}}\right)^{-12}\,,
\end{align}
which is regular at $R\to0$, $R\to \rho_H$ and $R\to\infty$.

Performing the conformal transformation \eqref{eq:metricregular} to the action for the cosmon \eqref{eq:cosmonaction} we obtain
\begin{align}\label{eq:regularcosmon}
S=\frac{1}{2}\int d^4 x\sqrt{-\bar g}\,\varphi^2\bar {\cal R}\,,
\end{align}
where we introduced a new regular scalar field by
\begin{align}\label{eq:regularcosmonfield}
\varphi\equiv{\rm e}^{\frac{\phi}{\sqrt{6}}}\,.
\end{align}
For the cosmon lump solution the radial dependence is given by
\begin{align}
\varphi(R)=\left|\frac{1- \rho_H/R}{1+ \rho_H/ R}\right|^{1/2}\,.
\end{align}

Interestingly, the resulting action after the transformation \eqref{eq:metricregular} is very similar to the action \eqref{eq:likegarriga} employed for the Hawking-Turok instanton in Sec.~\ref{sec:hawkingturok}. We are lacking the kinetic term for the scalar field like in Eq.~\eqref{eq:likegarriga} and this time there is no potential for the scalar field. Other possible frames with an action with a regular kinetic term as in Sec.~\ref{sec:hawkingturok} are discussed in App.~\ref{App:alternative}. The Einstein equations read \begin{align}
\varphi^2 \bar G_{\mu\nu}=\left(\bar\nabla_\mu\bar\nabla_\nu-\bar g_{\mu\nu}\bar\Box\right)\varphi^2\,,
\end{align}
and the scalar field equations yield
\begin{align}
 \varphi\bar{\cal R}=0\,.
\end{align}
Then, if $\varphi\neq 0$, we have that $\bar{\cal R}=0$ and $\bar \Box\varphi=0$ from the trace of Einstein equations. All quantities in these equations are regular everywhere. The energy momentum tensor of the scalar field must be regular since $\bar R_{\mu\nu}$ is regular. This implies that all curvature invariants are regular everywhere. For instance,
\begin{align}
\bar {\cal R}^2=0\,,\qquad \bar {\cal R}_{\mu\nu}\bar {\cal R}^{\mu\nu}=\frac{24\rho_H^2}{R^6}\left(1+\frac{\rho_H}{R}\right)^{-12}\,,\qquad\bar {\cal R}_{\alpha\beta\sigma\rho}\bar {\cal R}^{\alpha\beta\sigma\rho}=\frac{96\rho_H^2}{R^6}\left(1+\frac{\rho_H}{R}\right)^{-12}\,,
\end{align}
and the Gauss-Bonet term vanishes, \textit{i.e.}
\begin{align}
\bar {\cal G}=\bar {\cal R}^2-4\bar {\cal R}_{\mu\nu}\bar {\cal R}^{\mu\nu}+\bar {\cal R}_{\alpha\beta\sigma\rho}\bar {\cal R}^{\alpha\beta\sigma\rho}=0\,.
\end{align}

We could start with the action \eqref{eq:regularcosmon}. The solution of the corresponding field equations for $\bar g_{\mu\nu}$ and $\varphi$ corresponding to Eqs.~\eqref{eq:criticalcosmonmetric2} and \eqref{eq:metricregular} reads
\begin{align}\label{eq:regularsolution2}
d\bar s^2=-dt^2+f(\bar r)d\bar r^2+\bar r^2 d\Omega^2\,,\qquad \varphi(\bar r)=\left|\frac{\rho- \rho_H}{\rho+ 3\rho_H}\right|^{1/4}=\left|\frac{\bar r- 4\rho_H}{\bar r}\right|^{1/4}\,.
\end{align}
Here the coordinate $\bar r$ is related to $\rho$ by
\begin{align}
\bar r= \rho+3\rho_H\,,\qquad \rho=\bar r -3 \rho_H\,,
\end{align}
and one has
\begin{align}
f(\bar r)=\frac{\bar r}{\bar r-4\rho_H}\,.
\end{align}
For $\bar r\to \infty$ one recovers flat space and $\varphi\to 1$. In these coordinates the metric is singular at $\bar r=4\rho_H$, where $\varphi$ vanishes. In terms of the coordinate $R$ the metric is given by Eq.\eqref{eq:metriciso} and the solution for $\varphi$ is given by Eq.~\eqref{eq:regularcosmonfield}.

The action \eqref{eq:regularcosmon} is scale invariant under a canonical dilatation transformation of $\bar g_{\mu\nu}$ and $\varphi$. A natural coupling to matter preserves scale symmetry \cite{Wetterich:1987fm,Wetterich:2019qzx}. In this case the expectation value of the Higgs doublet is proportional to $\varphi$, and the same holds for the confinement scale in strong interactions and for all other mass scales in particle physics. After a Weyl scaling to the Einstein frame the coefficient of the curvature scalar is given by a fixed squared Planck mass $M_{\rm pl}^2=1$, and all particle masses are constant in the Einstein frame,
\begin{align}\label{eq:conformaleinstein}
g_{\mu\nu}=\varphi^{-2}\bar g_{\mu\nu}\,,\qquad S=\int d^4x\sqrt{-g}\left[\frac{1}{2}{\cal R}-3\partial_\mu\varphi\partial^\mu\varphi+...\right]\,,
\end{align}
where the dots stand for the matter contributions that can have only couplings to derivatives of $\varphi$. The positive kinetic term in the Einstein frame can be brought to canonical form by rescaling of $\varphi$. The conformal transformation \eqref{eq:conformaleinstein} is singular for $\varphi=0$. For this reason the regular solution \eqref{eq:regularsolution2} becomes singular in the Einstein frame. This produces the central singularity in the gray hole solution \eqref{eq:criticalcosmonmetric}. Obviously, this singularity is a field singularity, introduced only by a singular field transformation. Our example is a perfectly viable black-hole type solution without any physical singularity. The only question is if this solution is stable.

\subsection{Horndeski Black Holes\label{subsec:horndenskiblackholes}}

It is a natural extension to study black hole solutions in the Horndeski theory, since we are dealing with disformal transformations and disformal transformations have been shown to close the form of the Horndeski Lagrangian \cite{Bettoni}. We will work with two main representatives that are asymptotically flat for simplicity but the analysis could be easily extended to more general cases. First we will consider a space-like scalar field and later a time-like scalar field.

\subsubsection{Space-like scalar}
We consider the solution given in Ref.\cite{Babichev:2017guv}. The Horndeski action,
\begin{align}
S=\int d^4x \sqrt{-g}\left[ \left( \zeta+\beta \sqrt{\frac{X}{2}} \right) {\cal R}-\frac{\eta}{2}X-\frac{\beta}{\sqrt{2X}}\left(\left(\Box\phi\right)^2-\nabla_\mu\nabla_\nu\phi\nabla^\mu\nabla^\nu\phi\right) \right] \,,
\end{align}
admits a static spherically symmetric solution ($q=0$, $\phi=\psi$),
\begin{align}
\phi'=\pm\frac{\sqrt{2}\beta}{\eta r^2}\sqrt{f} \,, \qquad h=f^{-1}=1-\frac{\mu}{r}-\frac{\beta^2}{2\zeta\eta r^2}\,.
\end{align}
The Weyl tensor for this solution reads
\begin{align}
{\cal C}_{\alpha\beta\sigma\rho}{\cal C}^{\alpha\beta\sigma\rho}=\frac{12 \left(\beta^2 + r \zeta \eta \mu\right)^2}{r^8 \zeta^2 \eta^2}\,.
\end{align}
The main problem with this solution is that it is only valid up to the horizon since the scalar field becomes imaginary below the horizon. We can map this solution up to the horizon to a Minkowski frame, since the solution is completely regular at the horizon. There are two solutions to $h=0$
\begin{align}
r_{\pm}=\frac{\mu}{2}\left(1\pm\sqrt{1+\frac{2\beta^2}{\zeta\eta\mu^2}}\right) \,,
\end{align}
but for $\beta>0$ there is a single horizon at $r_+$. After integrating the scalar field we can write $r$ in terms of $\psi$ by
\begin{align}
r=r_+\frac{1+\tan^2\left[\frac{\phi-\phi_0}{\phi_A}\right]}{\tan^2\left[\frac{\phi-\phi_0}{\phi_A}\right]-\frac{r_+}{|r_-|}} \,,
\end{align}
where $\phi_A=2\sqrt{2}\beta/(\eta\sqrt{r_+|r_-|})$ and $\phi_0$ is an arbitrary constant. The transformation to Minkowski spacetime is given by 
\begin{align}
\omega^2=h^{-1}\,,\qquad D=\frac{r^4\eta^2}{2\beta^2}\left[h\left(1-\frac{1}{2}\frac{\partial \ln h}{\partial\ln r}\right)^2-1\right]\,.
\end{align}
Note that in this case we are mapping the horizon to infinity as $\bar r=h^{-1/2}r\to\infty$ when $r\to r_+$. In the new frame, the geometry and the field are regular, so that we found a regular frame.

\subsubsection{Time-like scalar}

It is interesting to use a time dependent scalar field as well, extending the range of application of these transformations. Here for simplicity we consider the stealth black hole solutions \cite{Babichev:2013cya}. For this solution, the action is given by
\begin{align}
S=\int d^4x \sqrt{-g} \Bigl( \zeta \, {\cal R}+\beta \, G^{\mu\nu}\partial_\mu\phi\partial_\nu\phi \Bigr)\,,
\end{align}
that corresponds to $G_4=\zeta-\frac{1}{2}\beta X$ in Horndeski theory. The stealth solution is given by
\begin{align}
h=f^{-1}=1-\frac{\mu}{r} \,, \qquad d\phi=q\,dt+q\sqrt{\frac{f-1}{h}}=q\,dt+\frac{q}{h}\sqrt{\frac{\mu}{r}}\,dr\,.
\end{align}
The scalar field is not ``seen'' by the geometry and it is regular, \textit{e.g.} $\phi=q\left(t+2\mu\left[\sqrt{r/\mu}-{\rm arctanh}(\sqrt{r/\mu})\right]\right)$. Note that the scalar field is proportional to the Gullstrand-Painlev{\'e} time. This solution has been found to be unstable in the asymptotically flat case but stable in asymptotically de Sitter space-times \cite{Babichev:2017lmw}. We will continue with the present solution for simplicity. As in Schwarzschild metric, the Weyl tensor is given by
\begin{align}
{\cal C}_{\alpha\beta\sigma\rho}{\cal C}^{\alpha\beta\sigma\rho}=\frac{12 \mu^2}{r^6} \,,
\end{align}
and it is singular at $r=0$. Nevertheless, the action vanishes and therefore is regular. This time the solution is regular all the way to the singularity and we can find a Minkowksi frame everywhere by solving Eq.~\eqref{eq:transformationrules} for a time-dependent field. The function $C$ has to obey the relation
\begin{align}
\left(1+\frac{\partial \ln \omega}{\partial\ln r}\right)^2=f+\frac{h\phi'^2}{q^2}\left(\omega^2 h-1\right)\,.
\end{align}
This differential equation is solved for
\begin{align}
\omega^2=\frac{5r}{4\mu}\,,
\qquad
D=\frac{1}{q^2}\left(1-\frac{9\mu}{5r}\right)\,,\qquad \bar r=\sqrt{\frac{5}{4\mu}}r^{3/2}\,.
\end{align}
Also the new time is given by
\begin{align}
d\bar t=dt-\left[\frac{1}{h}\sqrt{\frac{\mu}{r}}\left(\omega^2 h-1\right)\right]dr=dt_{GP}-\frac{5}{4}\sqrt{\frac{r}{\mu}}dr\,,
\end{align}
which is well behaved everywhere. We have found an exact field transformation which maps the stealth Schwarzschild to Minkowski space. Both the scalar field and the geometry are regular. 

The next question is how do we obtain a simple $r$ dependence with a time-like scalar field. The answer is that we need to consider a transformation involving second derivatives of the scalar field \cite{Jirousek:2018ago} or first derivatives of the metric \cite{Aoki:2018zcv}. For example, $\Box\phi=\frac{3}{2}q\mu^{1/2}r^{-3/2}$ and we can write our transformation in terms of $\Box\phi$ only. Here, we propose a new class of transformations with second derivatives and we leave in Appendix \ref{app:higherderivatives} how one may build invertible transformations using second derivatives.

In passing, we present a new form of the Schwarzschild metric which is an alternative to the Kerr-Schild \cite{Visser:2007fj} and the Gordon \cite{Liberati:2018osj} forms. The metric can be written in terms of derivatives of a scalar $\varphi$ as
\begin{align}
g^{\rm Sch}_{\mu\nu} = A(\bar r)\eta_{\mu\nu}+B(\bar r)\partial_\mu\varphi\partial_\nu\varphi \,,
\end{align}
or alternatively
\begin{align}
 d s^2 = g^{\rm Sch}_{\mu\nu} dx^\mu dx^\nu 
 = A(\bar r) \overline{\eta}_{\mu \nu} d\bar{x}^\mu d\bar{x}^\nu
 + B(\bar r)\left( \frac{\pa \varphi}{\pa \bar{x}^\mu} d\bar{x}^\mu \right)^2 \,,
\end{align}
where $\bar{\eta}_{\mu \nu}$ is the Minkowski metric in spherical coordinates ($\bar t$ and $\bar r$). The functions\footnote{Not to be confused with the functions $A$ and $B$ from Weinberg's notation for the spherically symmetric metric \cite{Weinberg:1972kfs}, $ds^2=-B(r)dt^2+A(r)dr^2+r^2d\Omega^2.$ } $A$, $B$ and $\varphi$ are given by
\begin{align}
A(\bar r)=\left(\frac{4}{5}\frac{\mu}{\bar r}\right)^{2/3} \,, \qquad B(\bar r)=\frac{9}{4}A(\bar r)-1 \,, \qquad \varphi=\bar t+\frac{\sqrt{5}}{3}\bar r\,.
\end{align}

\section{Conclusions\label{sec:conclusions}}

Singularities are unavoidably present in General Relativity under few assumptions \cite{Hawking2014,Borde:2001nh}. In particular, this statement assumes that matter satisfies the weak energy condition and that it is minimally coupled to gravity. In this paper, inspired by the work of Hawking and Turok where they proposed that a singular instanton with a regular action is physically acceptable \cite{Hawking:1998bn}, we study whether a given solution with a singularity whose total action is regular, admits a frame (after a metric transformation) where the geometry and the fields are regular. In a sense, we are breaking the assumption that matter is minimally coupled, since frame transformations typically induce non-minimal matter couplings. The requirement that the action is regular is a necessary condition for the solution to be regular but it is not sufficient. We allowed the metric transformation to be singular at the singular point since a regular metric transformation cannot remove any singularity. In this line, we showed in Sec.~\ref{sec:hawkingturok} that there is a regular frame for the Hawking-Turok instanton which interestingly coincides with the regular action derived in an embedding in one higher dimension \cite{Garriga:1998ri}. 

Motivated by the first result, we studied if a similar logic applies for static spherically symmetric black hole like systems with scalar hair (\textit{e.g.} with a non-trivial gradient of the scalar field). Since the Weyl tensor (which diverges at the singularity) transforms multiplicatively under conformal transformations we extended the analysis to general disformal transformations. We aim to find a disformally related metric where the Weyl tensor is regular. This is possible since disformal transformations stretch a privileged direction, which allows us to cancel any singular shear component near the singularity. We assume that the scalar field is a solution of the form $\phi=q\, t+\psi(r)$ (where $t$ is the coordinate time and $r$ the radial coordinate). In this case we find a scalar-field dependent disformal transformation of the metric such that in the new frame spacetime is Minkowski space, without any geometric singularity. For a convenient definition of the scalar field also the scalar field solution can be made regular in the new frame. Since the employed disformal transformation is singular, it is possible that in the new frame there is a matter-gravity coupling that becomes singular. This is not necessarily the case, however. 

The disformal transformation involves the scalar field. If the solution for this field before the transformation depends only on $r$ ($q=0$) we have shown that after a disformal transformation the metric is still static and spherically symmetric. In general, if the metric has a singularity behind a horizon, then a disformal transformation is only able to ``remove'' it if $q\neq 0$. Instead, if the metric presents a naked singularity and the scalar field solution is static ($q=0$), the disformal transformation permits to find a frame where the geometry is regular. 

To see whether the field content after the transformation is regular as well, we used four particular solutions with scalar hair. In the first example, we considered a singular solution of general relativity plus a massless scalar field known as cosmon lump (which corresponds to a singular Brans-Dicke type I solution). The geometry is rather close to a black hole geometry. However, it has no horizon. The behavior near the Schwarzschild radius is smoothened. We found that the naked singularity at the center can be mapped to a regular point in a Minkowksi metric. For a particular ``critical cosmon lump'' the singular solution in the Einstein frame can be mapped to a regular solution in the scaling frame by a conformal transformation. In the second and third examples, we considered two shift-symmetric Horndeski black holes with scalar hair, respectively a space-like and a time-like scalar field. In the case where the field is space-like the solution is only valid up to the horizon. We showed that the metric near the horizon can be mapped to a Minkowski metric. In the case of a time-like scalar field, the metric is Schwarzschild with a stealth scalar field. This case is particularly interesting because the metric can be mapped completely to a Minkowski metric if we allow the coefficients of the transformation to depend on second derivatives of the scalar field. In passing, we have given a new form to write the Schwarzschild metric. We have proposed a disformal invariant tensor in App.~\ref{app:disformalinvariant}.
 
Unfortunately, in most of these examples the action after the disformal transformation is rather involved and the presence of higher derivatives obscures the notion of a canonical field and the definition of a clear energy momentum tensor for the scalar field. Therefore, to conclude whether the field content is completely regular we would need a dedicated study of the resulting Lagrangian. An exception is the critical cosmon lump in Sec.~\ref{subsec:critical} where the singularity can be removed by a conformal transformation. The action in the scaling frame turns out to be very simple and resembles that of the Hawking-Turok's regular frame of Sec~\ref{sec:hawkingturok}. The action in the scaling frame is scale invariant provided the mass terms in the matter part depend on the scalar field appropriately. In this case one finds in the Einstein frame a realistic coupling of the scalar field to matter, with constant particle masses. The black hole type solutions to the field equations in the scaling frame show a completely regular geometry, with the scalar field vanishing at the center with a fractional power. All particle masses vanish at the center as well. The energy momentum tensor remains finite. From the point of view of singularities this is a completely regular solution without any physical singularities. The black hole singularity in the Einstein frame is only introduced by a singular field transformation. The critical cosmon lump may not be stable with respect to small perturbations--this issue needs further discussion.

In summary, we have shown that disformal transformations can significantly change the geometry of static spherically symmetric space-times. A singularity in the squared Weyl tensor squared is not straightforwardly related to a physical singularity, since a disformal transformation might yield a regular value for the squared Weyl tensor. Whether this yields a regular frame without physical singularities or it is simply a different interpretation of a physical singularity depends on the particular solution and is left for further study.

\section*{Acknowledgments}

G.D. would like to thank J.~Rubio for useful discussions and to propose to build a disformal invariant tensor. This work was partially supported by DFG Collaborative Research center SFB 1225 (ISOQUANT)(C.W. and G.D.). Some of the calculations were done with the \texttt{xCoba (xAct) Mathematica} package. 
A.N. is supported in part by a JST grant ``Establishing a Consortium for the Development of Human Resources in Science and Technology''.
The work of A.N. was partly supported by JSPS KAKENHI Grant No. JP19H01891.
This work is supported in part by MEXT grant Nos. 15H05888 and 15K21733.

\appendix

\section{Cosmon lumps in isotropic coordinates \label{app:cosmonlumps}}

In this appendix, we bring the Cosmon lumps solution into the isotropic form for a comparison with the literature. By the following coordinate transformation,
\begin{align}
\rho-\rho_H=R\left({1-\frac{R_s}{8\delta R}}\right)^2\,,
\end{align}
the metric in isotropic coordinates reads
\begin{align}\label{eq:isotropiccosmon}
ds^2=-h(R)dt^2+g(R)\left(dR^2+R^2d\Omega^2\right) \,,
\end{align}
where
\begin{align}
h(R)\equiv\left[\frac{1- R_s/(4\delta R)}{1+ R_s/(4\delta R)}\right]^{2\delta} \,, \qquad g(R)\equiv\left({1-\frac{R_s}{4\delta R}}\right)^{2(1-\delta)}\left({1+\frac{R_s}{4\delta R}}\right)^{2(1+\delta)} \,,
\end{align}
and
\begin{align}
\phi=\phi_0+\sqrt{2}\gamma\delta\ln\left[\frac{1- R_s/(4\delta R)}{1+ R_s/(4\delta R)}\right]\,.
\end{align}
This form is useful for a direct comparison with the literature (see Ref.~\cite{Faraoni:2016ozb} for a review). Also one can see the naked singularity is at $R=R_s/4\delta$ where space-time ends and this is why the limit $\gamma\to 0$ ($\delta\to 1$) does not contain the whole Schwarzschild space-time.
\subsection{Conformal transformation}
We can check whether the singularity may be removed by a simple conformal transformation, such as
\begin{align}\label{eq:confcosmon}
d\bar s^2&={\rm e}^{-\frac{\beta}{\delta}\frac{\sqrt{2}}{\gamma}(\phi-\phi_0)}ds^2=\left[\frac{1- R_s/(4\delta R)}{1+ R_s/(4\delta R)}\right]^{-2\beta}{ds^2}\nonumber\\&
=-\left[\frac{1- R_s/(4\delta R)}{1+ R_s/(4\delta R)}\right]^{2(\delta-\beta)}dt^2+\left({1-\frac{R_s}{4\delta R}}\right)^{2(1-\delta-\beta)}\left({1+\frac{R_s}{4\delta R}}\right)^{2(1+\delta+\beta)}\left(dR^2+R^2d\Omega^2\right) \,.
\end{align}
This metric corresponds to the Jordan frame Brans-Dicke type I solutions as in Ref.~\cite{Faraoni:2016ozb} which may be a naked singularity or a wormhole depending on the choice of $\beta$. In this case we have that the Ricci scalar and the Weyl tensor squared are respectively given by
\begin{align}
\bar{\cal R}=\frac{1}{R^2}\frac{R_s^2}{2\delta^2R^2}\left(1-\delta^2-3\beta^2\right)\left({1-\frac{R_s}{4\delta R}}\right)^{2(\delta+\beta-2)}\left({1+\frac{R_s}{4\delta R}}\right)^{-2(\delta+\beta+2)}
\end{align}
and
\begin{align}
\bar {\cal C}_{\alpha\beta\sigma\rho}\bar {\cal C}^{\alpha\beta\sigma\rho}=\frac{12}{R^4}\frac{R_s^2}{R^2}\left(1-\frac{R_s}{6\delta^2R}\left(1+2\delta^2\right)+\frac{R_s^2}{16\delta^2R^2}\right)\left({1-\frac{R_s}{4\delta R}}\right)^{4(\delta+\beta-2)}\left({1+\frac{R_s}{4\delta R}}\right)^{-4(\delta+\beta+2)}\,.
\end{align}
Note how the Weyl tensor squared is conformal invariant except for an overall conformal factor. We see that we cannot have a regular metric and regular curvature by any choice of $\beta$ as we cannot remove the term $\left({1-\frac{R_s}{4\delta R}}\right)$ from both the temporal and spatial components at the same time, except when $\delta=1/2=\beta$. 

\section{Critical cosmon lump alternative regular action \label{App:alternative}}
In Sec.\ref{subsec:critical} we found that the resulting action after the conformal transformation \eqref{eq:metricregular} is very similar to the results of Sec.~\ref{sec:hawkingturok}. Thus, we may try as well to find a regular frame which also has a regular kinetic term for the scalar field. We first do a conformal transformation given by
\begin{align}
\bar g_{\mu\nu}=\Omega^2 \tilde g_{\mu\nu}=\left(\frac{1+\varphi^2}{2}\right)^{2\alpha} \tilde g_{\mu\nu}\,,
\end{align}
which at $R\to \rho_H$ ($\varphi\to 0$) both metrics coincide up to a constant as $\Omega\to2^{-\alpha}$.
Then the resulting metric and action \eqref{eq:regularcosmon}  are respectively given by
\begin{align}
d\tilde s^2=-\left({1+\frac{\rho_H}{R}}\right)^{-2\alpha} dt^2+\left({1+\frac{\rho_H}{ R}}\right)^{4-2\alpha}\left(dR^2+R^2d\Omega^2\right)\,,
\end{align}
and
\begin{align}
S=\int d^4 x\sqrt{-\tilde g}\,\frac{1}{2}\,\varphi^2\left(\frac{1+\varphi^2}{2}\right)^{2\alpha}\tilde {\cal R}+6\varphi^2\alpha\left(\frac{1+\varphi^2}{2}\right)^{2\alpha-1}\left[1+\alpha\frac{\varphi^2}{1+\varphi^2}\right]\tilde X\,,
\end{align}
where $\tilde X=\tilde\nabla_\mu\varphi\tilde\nabla^\mu\varphi$. The Ricci scalar is given by
\begin{align}
\tilde {\cal R}=\frac{12}{R^2}\alpha\left(1-\alpha\right)\frac{\rho_H}{R^2}\left({1+\frac{\rho_H}{R}}\right)^{2\alpha-6} \,,
\end{align}
and the Weyl tensor squared by
\begin{align}
\tilde {\cal C}_{\alpha\beta\sigma\rho}\tilde {\cal C}^{\alpha\beta\sigma\rho}=\frac{48}{R^4}\frac{\rho_H^2}{R^2}\left({1+\frac{\rho_H}{R}}\right)^{4\alpha-12}\,.
\end{align}

We first notice that everything is regular at $R=\rho_H$ (where the singularity was before) and if $\alpha<0$ and $|\alpha|<2$ then the kinetic term of the field has the correct sign everywhere as $0<\varphi<1$. However, since we are in the Jordan frame the sign of the kinetic term does not necessarily imply a ghost \cite{maedabook}. To be more concrete, we can check that near the previously naked singularity at $R\sim \rho_H$ where $\varphi\sim 0$ the action becomes for $\alpha<0$
\begin{align}
S\sim\int d^4 x\sqrt{-\tilde g}\left(\frac{1}{2}\,\tilde\varphi^2\tilde {\cal R}-12\times2^{2\alpha}|\alpha|\tilde \varphi^2\tilde\nabla_\mu\tilde\varphi\tilde\nabla^\mu\tilde\varphi\right)\,,
\end{align}
where $\tilde\varphi\equiv 2^{-\alpha}\varphi$. In this new frame, the metric, the Ricci curvature, the Weyl tensor squared, the field and its first two derivatives are regular everywhere. We thus have found a regular frame for the naked singularity with $\delta=1/2$. This result gives support to the conclusion that the singular cosmon solution is a field singularity and can be removed by a disformal transformation.

\section{Disformal transformation formulas \label{app:disformulas}}
The Horndeski action is given by
\begin{align}
S=\int d^4x \sqrt{-g}\sum_i{\cal L}_i\,,
\end{align}
where
\begin{align}
	{\cal L}_2&=G_2(\phi,X) \,, \\
	{\cal L}_3&=G_3(\phi,X)\Box\phi \,,\\
	{\cal L}_4&=G_4(\phi,X){\cal R}-2G_{4,X}\left[\left(\Box\phi\right)^2-\nabla_\mu\nabla_\nu\phi\nabla^\mu\nabla^\nu\phi\right] \,, \\
	{\cal L}_5 &=G_5(\phi,X)G^{\mu\nu}\nabla_\mu\nabla_\nu\phi -2G_{5,X}\left[\left(\Box\phi\right)^3-3\Box\phi\nabla_\mu\nabla_\nu\phi\nabla^\mu\nabla^\nu\phi+2\nabla_\mu\nabla_\nu\phi\nabla^\mu\nabla^\alpha\phi\nabla_\alpha\nabla_\mu\phi\right] \,.
\end{align}
Here we have defined $X=\nabla^\mu\nabla_\mu\phi$, $\Box\phi\equiv\nabla^\mu\nabla_\mu\phi$ and $G^{\mu\nu}$ is the Einstein tensor. Here we give the transformation formulas for a subset of Horndeski action with $G_2=G_2(X)$, $G_3=G_5=0$ and $G_4=G_4(X)$. Under
\begin{align}\label{eq:disfinv}
g_{\mu\nu}= A(\phi) \bar g_{\mu\nu}+B(\phi)\nabla_\mu\phi\nabla_\nu\phi\,.
\end{align}
 we find
\begin{align}
\bar G_2 = A^2 {\cal G}_2 + A_\phi \bar X{\cal G}_3-2 \bar X {\cal G}_4 A_{\phi\phi} +\frac{3}{2}\frac{A_\phi^2}{A}\bar X{\cal G}_4\,,
\end{align}
\begin{align}
\bar G_3= A {\cal G}_3+2A_\phi\left(
{\cal G}_4+\bar X {\cal G}_{4,\bar X}\right)\,,
\end{align}
and
\begin{align}
\bar G_4= A {\cal G}_4 \,.
\end{align}
Here
\begin{align}
{\cal G}_2 &= \sqrt{1+\frac{B}{A}\bar X}  G_2 -\frac{\bar X}{2A^2}\int d\bar X \frac{\left(G_4-2\bar X G_{4\bar X}\right)}{\sqrt{1+\frac{B}{A}\bar X}}\left[B_{\phi\phi}-\frac{B_\phi^2}{2A}\frac{\bar X}{1+\frac{B}{A}\bar X}\right]\,, \\
{\cal G}_3 &= -\frac{B_\phi\bar X}{A} \frac{\left(G_4-2\bar X G_{4\bar X}\right)}{\sqrt{1+\frac{B}{A}\bar X}}+\frac{B_\phi\bar X}{2A^2}\int d\bar X \frac{\left(G_4-2\bar X G_{4\bar X}\right)}{\sqrt{1+\frac{B}{A}\bar X}}\,, \\
{\cal G}_4 &= \sqrt{1+\frac{B}{A}\bar X}  { G}_4\,.
\end{align}
We performed first a disformal transformation yielding the coefficients ${\cal G}_i$ and later a conformal transformation yielding the coefficients $\bar G_i$. Note that the arguments of the barred functions are now to be intended as functions of the matter frame variables, \textit{e.g.} $X= \bar X/(A+B\bar X)$.

\subsection{Cosmon lumps\label{eq:cosmondisfaction}}

Here we present the explicit expression for the resulting action after a disformal transformation to Minkowski spacetime in Sec.~\ref{subsec:cosmonlumps}. The functions of the corresponding disformal transformation for the cosmon lumps Eq.~\eqref{eq:disfinv} are given by
\begin{align}
A=\omega^{-2}=h\qquad{\rm and}\qquad B=-D=\left[\left(1-\frac{1}{2}\frac{\partial\ln h}{\partial \ln r}\right)^{2}-f\right]\left(\frac{\partial\phi}{\partial r}\right)^{-2}\,,
\end{align}
where $C(\phi)$ and $D(\phi)$ are given by Eq.~\eqref{eq:distranscosmon}.
The resulting action after the transformation reads
\begin{align}
S=\int d^4x \sqrt{-g}\frac{1}{2}A^2\sqrt{1+B\bar X/A}\Bigg\{\bar{\cal R}&-\frac{\bar X}{1+B\bar X/A}\left[1-\frac{3}{2}\left(\frac{\partial\ln A}{\partial \phi}\right)^2\right]-\frac{B/A}{1+B\bar X/A}\left[\left(\bar\Box\phi\right)^2-\bar\nabla_\mu\bar\nabla_\nu\phi\bar\nabla^\mu\bar\nabla^\nu\phi\right]\nonumber\\&
-\frac{B/A}{1+B\bar X/A}\left[\left(\frac{\partial\ln A}{\partial \phi}\right)+\left(\frac{\partial\ln B}{\partial \phi}\right)\right]\left[\bar X\bar\Box\phi-\frac{1}{2}\bar\nabla_\nu\phi\bar\nabla^\nu\bar X\right]\Bigg\}\,.
\end{align}
In this frame, the geometry is indeed completely regular as it is Minkowski spacetime. Although we know that the barred frame action vanishes, as the value of the unbarred action \eqref{eq:cosmonaction} is zero, it is difficult to argue that there is a field redefinition which makes the field content regular as well. The reason is that in the presence of higher derivatives of the scalar field, there is no clear notion of energy momentum tensor as the scalar field mixes with gravity. We may find a redefinition of the scalar field which makes the value of the field and its derivatives regular but it does not guarantee that each term in the action will be regular as well.

\section{Metric transformation with higher derivatives \label{app:higherderivatives}}

In this appendix we develop how to build invertible metric transformations with second derivatives. We aim to build a quantity which is invariant or rescales with the metric transformation and, thus, the invertibility of the transformation is reduced to an algebraic equation. This is similar to Ref.~\cite{Jirousek:2018ago} where the scalar field had a Weyl weight to compensate the conformal transformations. Here we will add an additional term that compensate the transformation of $\Box\phi$. We will do so separately for a conformal and a disformal transformation. 

First, under a conformal transformation $\bar g_{\mu\nu}=\lambda \, g_{\mu\nu}$ the following object,
	\begin{align}
	Q_C\equiv\Box\phi + \nabla^\mu\phi\nabla_\mu \ln X\,,
	\end{align}
just rescales to $\bar Q_C=Q_C/\lambda$. If $\lambda=\lambda(Q_C)$ then the transformation is invertible once we find $Q_C(\bar Q_C)$. Second, under a disformal transformation $\tilde g_{\mu\nu}=g_{\mu\nu}+D\partial_\mu\phi\partial_\nu\phi$ the following object,
	\begin{align}
	Q_D\equiv\frac{1}{X}\left(\Box\phi -\frac{1}{2} \nabla^\mu\phi\nabla_\mu \ln X\right)\,,
	\end{align}
	is invariant, $\tilde Q_D=Q_D$. So that a disformal transformation with $D=D(Q_D)$ is invertible. We see that in the uniform-$\phi$ slicing they are given by
	\begin{align}
	Q_C=\epsilon \left(3{\cal L}^2_n\phi+K{\cal L}_n\phi\right) \,, \qquad
	Q_D=\epsilon K/{\cal L}_n\phi\,,
	\end{align}
	where ${\cal L}_n\phi=n^\mu\nabla_\mu\phi$ and $\epsilon\equiv n^\mu n_\mu$. We can see that while $Q_D$ does not involve second time derivatives, $Q_C$ does. In both cases that leads to $Q_C(r)$ and $Q_D(r)$ in the stealth black hole space-time, Sec.~\ref{subsec:horndenskiblackholes}.

\section{Disformal Invariant tensor\label{app:disformalinvariant}}
In this appendix we derive an object invariant under a pure general disformal transformation given by
\begin{align}
\bar g_{ab}=g_{ab}+ D(x^\mu)\partial_a\phi\partial_b\phi \,,
\end{align}
with $D(x^\mu)$ an arbitrary function of the coordinates $x^\mu$. The inverse metric is given by
\begin{align}
\bar{g}^{ab}=g^{ab}-\frac{D}{1+DX}\nabla^a\phi\nabla^b\phi\,,
\end{align}
where $X\equiv\nabla_a\phi\nabla^a\phi$. The change in the connection is given by
\begin{align}
{\cal D}^c_{ab}=\bar{\Gamma}^c_{ab}-{\Gamma}^c_{ab}=\bar g^{ce}\left(\nabla_{(a}\bar g_{b)e}-\frac{1}{2}\nabla_e \bar g_{ab}\right)\,,
\end{align}
and its explicit expression in terms of the field $\phi$ reads
\begin{align}
{\cal D}^c_{ab}&=\frac{D}{1+DX}\n^c\phi\left(\nabla_a\nabla_b\phi+\nabla_{(a}\phi\nabla_{b)}\ln D+\frac{1}{2}\nabla^d\phi\nabla_d D\nabla_a\phi\nabla_b\phi\right)
	-\frac{1}{2}\nabla^c D\nabla_a\phi\nabla_b\phi\,.
\end{align}
The transformation of the Riemann tensor is thus
\begin{align}
\bar{\cal R}^a\,_{bcd}={\cal R}^a\,_{bcd}+2\nabla_{[c}{\cal D}^a_{d]b}+2{\cal D}^a_{e[c}{\cal D}^e_{d]b}\,.
\end{align}
It is simpler to compute the transformation of the Riemann tensor with lower indexes which reads
\begin{align}
\bar{\cal R}_{abcd}=\bar g_{ae}{\cal R}^e\,_{bcd}={\cal R}_{abcd}+D\nabla_a\phi\left(\nabla_d\nabla_c\nabla_b\phi-\nabla_c\nabla_d\nabla_b\phi\right)+2\bar g_{ae}\left(\nabla_{[c}{\cal D}^a_{d]b}+{\cal D}^a_{e[c}{\cal D}^e_{d]b}\right) \,,
\end{align}
where we have used that
\begin{align}
{\cal R}^e\,_{bcd}\nabla_e\phi\nabla_a\phi=
\nabla_a\phi\left(\nabla_d\nabla_c\nabla_b\phi-\nabla_c\nabla_d\nabla_b\phi\right)\,.
\end{align}
After a long algebra we obtain that
\begin{align}
\bar{\cal R}_{abcd}=&{\cal R}_{abcd}+2\frac{D}{1+DX}\nabla_a\nabla_{[c}\phi\nabla_{d]}\nabla_b\phi+\frac{X}{1+DX}\nabla_{[a}\phi\nabla_{b]}D\nabla_{[c}\phi\nabla_{d]}D+2\nabla_{[a}\phi\nabla_{b]}\nabla_{[c}D\nabla_{d]}\phi\\&
-2\frac{D}{1+DX}\nabla_e\phi\nabla^e D\nabla_{[a}\phi\nabla_{b]}\nabla_{[c}\phi\nabla_{d]}\phi-2\frac{1}{1+DX}\left(\nabla_{[a}\phi\nabla_{b]}\nabla_{[c}\phi\nabla_{d]}D+\nabla_{[c}\phi\nabla_{d]}\nabla_{[a}\phi\nabla_{b]}D\right)\,.
\end{align}

We found that the following object,
\begin{align}
{\cal D}_{abcd}\equiv &{\cal R}_{abcd}+\frac{2}{X}\nabla_{c}\nabla_{[a}\phi\nabla_{b]}\nabla_{d}\phi+\frac{2}{X^2}
\nabla_{[a}\phi\nabla_{b]}\nabla_{[c}X\nabla_{d]}\phi
+\frac{4}{X^3}\nabla_{[a}\phi\nabla_{b]}X\nabla_{[c}\phi\nabla_{d]}X\nonumber\\&-\frac{2\nabla_\mu\phi\nabla^\mu X}{X^3}
\nabla_{[a}\phi\nabla_{b]}\nabla_{[c}\phi\nabla_{d]}\phi+\frac{C(\phi)}{X}
\nabla_{[a}\phi\nabla_{b]}\nabla_{[c}\phi\nabla_{d]}\phi \,,
\end{align}
is disformal invariant under a general disformal factor $D$. This tensor is in general not invariant under a conformal transformation. In the above equation the antisymmetric bracket is normalized and the last term is an arbitrary term that is disformal invariant by itself and that appears after a field redefinition $\phi=f(\varphi)$. These invariance holds for any disformal factor and for any field redefinition up to a different $C$. 

From the tensor ${\cal D}_{abcd}$ we can build various disformal invariant scalars with the projector
\begin{align}
P^{ab}=g^{ab}-\frac{1}{X}\nabla^a\phi\nabla^b\phi \,,
\end{align}
which is disformal invariant. For example the projection of the disformal invariant tensor given by
\begin{align}
P^{ac}P^{bd}{\cal D}_{abcd}=P^{ac}P^{bd}\left({\cal R}_{abcd}+\frac{2}{X}\nabla_{c}\nabla_{[a}\phi\nabla_{b]}\nabla_{d}\phi\right) \,,
\end{align}
is disformal invariant as well as invariant under field redefinitions. The meaning of this tensor is not yet clear to us. 

\subsection{Static spherically symmetric metric}
For a general static spherically symmetric metric given by
\begin{align}
ds^2=-h(r)dt^2+f(r)dr^2+r^2d\Omega^2 \,,
\end{align}
and a scalar field with profile
\begin{align}
\phi(t,r)=qt+\psi(r)\,,
\end{align}
we find that
\begin{align}
P^{ac}P^{bd}{\cal D}_{abcd}=\frac{2}{r^2\left(q^2f-h\psi'^2\right)^2}\Big[q^4\left(f^2-f+rf'\right)-q^2\psi'\left(2hf+\left[rh'-h\right]\psi'+2rh\psi''\right)+h^2\psi'^4\Big]\,.
\end{align}
When $q\to0$ we have that
\begin{align}
P^{ac}P^{bd}{\cal D}_{abcd}=\frac{2}{r^2} \,,
\end{align}
and coincides with the value in Minkowski spacetime.

\subsection{Cosmon lumps}
If we evaluate it for the cosmon solution of Sec.~\ref{subsec:cosmonlumps} in the isotropic coordinates \eqref{eq:isotropiccosmon}, we find that 
\begin{align}
P^{ac}P^{bd}{\cal D}_{abcd}=\frac{2}{R^2} \left({1-\frac{R_s}{4\delta R}}\right)^{-2(1-\delta)}\left({1+\frac{R_s}{4\delta R}}\right)^{-2(1+\delta)}=\frac{2}{g(R)R^2}\,.
\end{align}
This result is divergent for $\delta<1$. However, the disformal invariant tensor is not invariant under a conformal transformation. Let us use instead the metric Eq.~\eqref{eq:confcosmon}, which is the original cosmon metric \eqref{eq:isotropiccosmon} after a conformal transformation. Then, we find that 
\begin{align}
P^{ac}P^{bd}{\cal D}_{abcd}\Big|_{g}=\frac{2}{R^2} \left({1-\frac{R_s}{4\delta R}}\right)^{-2(1-\delta-\beta)}\left({1+\frac{R_s}{4\delta R}}\right)^{-2(1+\delta+\beta)}\,.
\end{align}
We see that if $\beta>1-\delta$, the disformal invariant tensor is regular. To be more specific let us choose $\beta=\delta$, which corresponds to the Minkowski spacetime barred metric studied in Sec.~\ref{subsec:cosmonlumps}. The metric in this case is given by
\begin{align}\label{eq:confcosmon}
d\bar s^2&=-dt^2+\left({1-\frac{R_s}{4\delta R}}\right)^{2-4\delta}\left({1+\frac{R_s}{4\delta R}}\right)^{2+4\delta}\left(dR^2+R^2d\Omega^2\right) \,.
\end{align}
The corresponding disformal invariant tensor is thus given by
\begin{align}
P^{ac}P^{bd}{\cal D}_{abcd}\Big|_{\bar g}=\frac{2}{R^2} \left({1-\frac{R_s}{4\delta R}}\right)^{-2+4\delta}\left({1+\frac{R_s}{4\delta R}}\right)^{-2-4\delta}=\frac{2}{\bar r^2}\,,
\end{align}
where 
\begin{align}
\bar r^2\equiv R^2\left({1-\frac{R_s}{4\delta R}}\right)^{2-4\delta}\left({1+\frac{R_s}{4\delta R}}\right)^{2+4\delta} \,, 
\end{align}
and corresponds to the radial coordinate in the resulting Minkowski spacetime, \textit{i.e.} Eq.~\eqref{eq:sphericallysymmetricmetric2} with $\bar h=\bar f=1$. If we evaluate the value of the disformal invariant tensor in Minkowski spacetime we indeed find that
\begin{align}
P^{ac}P^{bd}{\cal D}_{abcd}\Big|_{\rm Mink.}=\frac{2}{\bar r^2}=P^{ac}P^{bd}{\cal D}_{abcd}\Big|_{\bar g}\,,
\end{align}
is disformal invariant. We also checked that it is also invariant under field redefinitions. The present divergence at $\bar r=0$ corresponds to the typical divergence in spherical coordinates since the scalar field $\phi$ is a function of $\bar r$ only. Although we need to check more examples to conclude any physical significance of the disformal invariant tensor, we believe that it may yield information on whether a metric can be made regular by a disformal transformation.
\subsection{Stealth black hole}
For the stealth black hole solution of Sec.~\ref{subsec:horndenskiblackholes} we find that
\begin{align}
P^{ac}P^{bd}{\cal D}_{abcd}\Big|_{\rm stealth}=0\,.
\end{align}
However, since we know that the disformal invariant tensor is not conformal invariant we should compute the value for the intermediate metric
\begin{align}
ds_{\rm stealth}^2=\omega^2 ds^2_{\rm C} \qquad {\rm with}\qquad \omega^2=\frac{5r}{4\mu}\,.
\end{align}
We have that
\begin{align}
P^{ac}P^{bd}{\cal D}_{abcd}\Big|_{\rm stealth,C}=-\frac{2\mu}{r^3}=-\frac{5}{2\bar r^2}=P^{ac}P^{bd}{\cal D}_{abcd}\Big|_{\rm stealth,Mink.}\,.
\end{align}
Again the object is disformal invariant and coincides with the value for Minkowski spacetime. It is not clear though how to conclude that the metric can be brought to Minkowski spacetime without knowing the correct disformal transformation beforehand.

\bibliography{dis_sing.bib}

\end{document}